\documentclass[intlimits,twoside,a4paper]{article}

\usepackage[cp1251]{inputenc}

\usepackage[eqsecnum]{cmpj3}
\usepackage{graphicx}

\usepackage{multirow}

\articletype{Review}

\issue{2026}{29}{1}{13802}
\doinumber{10.5488/CMP.29.13802}

\title[Theory of hemicellulose and lignin in solution]%
{Multiscale theory, modelling, and simulation of hemicellulose and lignin in solution\thanks{This paper is dedicated to the memory of Stefan Soko{\l}owski, Professor at Maria Curie-Sk{\l}ododowska University in Lublin, Poland, an extraordinary scientist in microscopic theory and computer simulations of fluids and solutions.}}

\author[A. Kovalenko]{A. Kovalenko\orcid{0000-0001-5033-4314}\thanks{Email: \email{akovalenko@smmodeling.com}}}
\address{ Software for Multiscale Modelling, Inc., Edmonton, Alberta, Canada T6E 5J5 }

\Keywords{cellulose nanocrystals, hemicellulose, lignin, molecular solvation theory, solvation thermodynamics, molecular simulations }

\date{Received 24 November 2025; revised 11 January 2026; accepted 14 January 2026; published 30 March 2026}
\begin{document}

\maketitle

\begin{abstract}
%\footnote{This paper is dedicated to the memory of Stefan Soko{\l}owski, Professor at Maria Curie-Sk{\l}ododowska University in Lublin, Poland, an extraordinary scientist in microscopic theory and computer simulations of fluids and solutions.}
This review examines multiscale modelling approaches for cellulose nanocrystals (CNCs) and lignocellulosic plant cell walls, with a focus on hemicellulose and lignin interactions in aqueous environments. The three-dimensional reference interaction site model with the Kovalenko–Hirata closure (3D-RISM-KH) is highlighted as a powerful molecular solvation theory applied in nanochemistry and biomolecular simulations. The method has been successfully employed to investigate hemicellulose hydrogels, the influence of hemicellulose composition on nanoscale forces in primary cell walls, and lignin-lignin and lignin-hemicellulose interactions. Findings indicate that these interactions are predominantly hydrophobic and entropy-driven, arising from water exclusion effects. Insights gained through this modeling framework deepen the understanding of molecular-scale forces in plant cell walls and inform strategies for biomass valorization, including genetic engineering and pretreatment technologies aimed at enhancing cellulose extraction and utilization.

\printkeywords

\end{abstract}

\setcounter{equation}{0}

\section{Introduction}

Molecular interactions in nanochemical and biomolecular systems differ considerably from those in bulk phases due to the effects of size, shape, and chemical composition. Understanding these interactions is essential for designing advanced materials and improving bioprocessing strategies. Various theoretical models have been developed to describe solvation and interaction processes in liquid media, ranging from continuum solvation frameworks to quasichemical approaches. However, these models often face limitations in system size and accuracy for complex biological and polymeric structures \cite{1,2,3,4,5,6,7,8,9}.

In the context of sustainable technologies, a major challenge lies in overcoming the intrinsic recalcitrance of plant biomass \cite{10,11}. Lignocellulosic materials possess a highly ordered and resistant architecture, where cellulose is closely associated with hemicellulose and lignin. These noncellulosic polymers act as protective matrices that restrict enzymatic access to cellulose, thereby limiting efficient deconstruction \cite{12,13}. Environmentally friendly pretreatment methods seek to fractionate lignocellulose into its major components while minimizing the energy input and chemical waste \cite{14}. Emerging catalytic processes such as “lignin-first” biorefining have shown promise in selectively removing lignin while preserving carbohydrate fractions, enabling the generation of high-value bioproducts \cite{15}. 

{
Multiscale frameworks combining molecular dynamics (MD), kinetic modelling, and machine learning are emerging to overcome the significant challenge of linking atomistic MD mechanisms to larger-scale process predictions \cite{16}. Much attention has been paid to using density functional theory (DFT) and MD to understand the solubilization and conversion processes of the cellulose, hemicellulose, and lignin components of plant biomass to produce high value-added chemicals and fuels \cite{17}. DFT is often used with model compounds (e.g., xylose for hemicellulose, guaiacyl dimers for lignin) because the full polymers are too complex for direct quantum treatment \cite{17}. Simulations of model lignocellulosic mixtures of cellulose, hemicellulose, and lignin using MD and DFT showed that intercomponent interactions influence the decomposition dynamics, that is hydrogen bonding networks affect accessibility and reaction initiation \cite{18}. DFT shows how solvent molecules or catalysts change the activation barriers, in particular, how energy barriers for certain lignin decomposition pathways are lowered by water/methanol \cite{19}. Studies of base-catalyzed $\beta$-O-4 cleavage provide insight into selective depolymerization mechanisms relevant to chemical valorization \cite{20}. DFT work on lignin dissolution in ionic liquids reveals hydrogen bonding interactions and structural effects relevant to pretreatment \cite{21}. DFT analysis of xylose, xylobiose, xylan decomposition reveals preferred pathways to furans and small carbonyl compounds~\cite{22}. For hemicellulose, DFT elucidates initial bond cleavages and subsequent fragmentation sequences which are valuable for kinetic modelling and understanding the product formation \cite{22}. MD studies of deep eutectic solvent (DES) interactions with hybrid systems of lignin and hemicellulose helped explain dissolution/pretreatment efficiency, especially via hydrogen bond analyses and interaction energies \cite{23}. 
}

{
The reference interaction site model (RISM) \cite{24,25,26} complemented by a variety of closure relations~\cite{3,4,5,6,7,8,9,24,25,26,27,28,29,30,31,32,33,34,35,36,37,38,39} provides an alternative capable of delivering molecular-level insights into solvation phenomena with computational efficiency. Its three-dimensional extension (3D-RISM) \cite{24,25} calculates solvent site distributions around solute molecules, enabling a spatially resolved description of solvation structure. Among the closure relations, the Kovalenko–Hirata (KH) approximation [28] has proven especially effective across diverse molecular systems, ranging from small molecules \cite{40} and biomacromolecules \cite{41} to nanomaterials \cite{42,43,44,45}, drug binding \cite{46,47}, ionic liquids \cite{48}, protein folding \cite{49}, and molecular recognition processes \cite{50,51}. Moreover, integration of 3D-RISM-KH with Density Functional Theory (DFT) \cite{52} has extended its applicability to electronic structure calculations.
}

By applying 3D-RISM-KH modelling, significant advances have been made in clarifying intermolecular interactions within plant cell walls {that were studied earlier in \cite{53,54}}. {These previous studies included} enzymatic deconstruction pathways, supramolecular polymer self-assembly \cite{55}, and cooperative interactions within polysaccharides \cite{56}. Such {previous} modelling has been instrumental in elucidating molecular recognition, carbohydrate–$\pi$, and $\pi$–$\pi$ stacking interactions, which underlie biological signaling and structural stability \cite{57,58}. Of particular interest, glucuronoarabinoxylan hemicellulose, which is a major constituent of biofuel feedstocks such as corn and sugarcane has been studied for its role in modulating cell wall mechanics. Genetic modifications altering hemicellulose composition have demonstrated a potential for improving cellulose accessibility without impairing plant growth \cite{13}.

Recent 3D-RISM-KH investigations have provided detailed insights into CNC-hemicellulose interactions, showing that carboxylate groups significantly contribute to the reinforcement of primary cell walls~\cite{59}. Furthermore, studies have confirmed that lignin-lignin and lignin-hemicellulose associations are largely hydrophobic and entropy-driven \cite{60}. These interactions are further stabilized by lignin methoxy groups, which strengthen the supramolecular matrix and thereby contribute to the persistence of biomass recalcitrance.

Collectively, multiscale modelling approaches such as 3D-RISM-KH offer a comprehensive understanding of the molecular determinants of cell wall integrity. This knowledge supports the development of improved pretreatment processes, genetic engineering strategies, and the design of bio-based materials with tailored mechanical properties.

\section{Molecular theory of solvation for nanochemical and biomolecular systems in electrolyte solutions}

\subsection{Integral equations}

The 3D-RISM-KH molecular theory of solvation \cite{61,62,63,64,65,66,67} is based on the probability density function $g_\gamma({\bf{ r}})$, which describes the spatial likelihood of locating solvent site $\gamma$ around a solute of arbitrary geometry. In solution, $g_\gamma({\bf r}) > 1$ indicates enhanced density regions, while $g_\gamma({\bf r}) < 1$  denotes depletion, converging to unity in the bulk. This distribution is determined by solving the 3D-RISM integral equation 
\begin{equation}
	h_\gamma({\bf r}) = \sum_\alpha \int \rd{\bf r^\prime} g_\alpha({\bf r-r^\prime}) \chi_{\alpha\gamma}(r^\prime) , 
	\label{eqn:3D-RISM}
\end{equation}
complemented with the 3D-KH closure approximation
\begin{equation}
	\label{eqn:3D-KH-closure}
	\begin{split}
		&  g_\gamma({\bf r}) = \begin{cases}
			\exp\left( d_\gamma({\bf r}) \right) & \text{for } g_{\alpha\gamma}({\bf r}) \leqslant 1 ,\\
			1 + d_\gamma({\bf r}) & \text{for } g_\gamma({\bf r}) > 1 ,
		\end{cases}  \\
		&  d_\gamma({\bf r}) = - u_\gamma({\bf r}) / \left(k_{\rm B}T\right) + h_\gamma({\bf r}) - c_\gamma({\bf r}),
	\end{split}
\end{equation}
which are solved for the 3D site total correlation function $h_\gamma({\bf r})$ and the 3D site direct correlation function $c_\gamma({\bf r})$ that follows the asymptotic behavior $c_\gamma({\bf r}) \sim - u_\gamma({\bf r}) / (k_{\rm B}T)$ related to the 3D solute-solvent interaction potential $u_\gamma({\bf r})$. The 3D-KH closure effectively combines the Mean Spherical Approximation (MSA) and Hypernetted Chain (HNC) approximation, applying MSA in density-enhanced regions and HNC in density-depleted areas. Unlike the 3D-RISM-HNC model, which may overestimate associative interactions and introduce numerical instability, 3D-RISM-KH demonstrates robustness in electrolyte and strongly interacting systems.

The solvent site-site susceptibility function 
\begin{equation}
	\chi_{\alpha\gamma}(r) = \omega_{\alpha\gamma}(r) + \rho_\alpha h_{\alpha\gamma}(r) 
	\label{eqn:chi}
\end{equation}
comprises the intramolecular matrix describing the geometry of solvent molecules and the site-site total correlation function $h_{\alpha\gamma}(r)$ of bulk solvent at number site density $\rho_\alpha$. These terms account for solvent molecular geometry and intermolecular correlations. The site-site total correlation functions are derived from the dielectrically consistent RISM (DRISM) theory of Perkyns and Pettitt \cite{68} incorporating dielectric bridge corrections to ensure consistency with macroscopic dielectric properties, which is complemented with the site-site KH closure \cite{25}. Such refinements make the 3D-RISM-KH theory broadly applicable to strongly associating species across varying temperatures, pressures, and solution compositions.

\subsection{Solvation free energy and potential of mean force}

Within the 3D-RISM-KH framework, solvation free energy of solute ``u'' in solvent ``v'' is expressed as an exact differential 
\begin{equation}
	\mu_{\rm u} = \epsilon_{\rm u} - TS_\text{V}
	\label{eqn:mu-solv}
\end{equation}
derived via Kirkwood’s thermodynamic integration as
\begin{subequations}  \label{eqn:mu-excess}
	\begin{align}
		\mu_{\rm u} & = \sum_\gamma \int {\rm d}{\bf r}\, \Phi_\gamma(\bf r) ,  \label{eqn:mu-excess-a} \\
		\Phi_\gamma(\bf r) & = \rho_\gamma k_{\rm B}T \left( \frac{1}{2} h^2_\gamma({\bf r})\Theta\left(-h_\gamma({\bf r})\right) - c_\gamma({\bf r}) - \frac{1}{2} h_\gamma({\bf r}) c_\gamma({\bf r}) \right) , \label{eqn:mu-excess-b}
	\end{align}
\end{subequations}
where $\Phi_\gamma(\bf r)$ denotes the 3D solvation free energy density, and $\Theta(x)$ is the Heaviside step function.
 
The solvation free energy is decomposed into the solute-solvent and solvent-solvent enthalpic contributions $\epsilon^{\rm uv}$ and $\epsilon^{\rm vv}$, and the entropic term $S_V$, 
\begin{subequations}	 \label{eqn:mu-energy-entropy}
	\begin{align}
	\mu_{\rm u} = \epsilon^{\rm uv} + \epsilon^{\rm vv} - TS_{\rm V},  \label{eqn:mu-energy-entropy-a} \\ 
		S_V = - \frac{1}{T} \left( \frac{\partial\mu}{\partial T} \right)_V,  \label{eqn:mu-energy-entropy-b} 
	\end{align}
\end{subequations}
while the partial molar volume is obtained using Kirkwood-Buff theory \cite{69} adapted for the 3D-RISM formalism as \cite{70,71} 
\begin{equation}
	\bar{V} = k_{\rm B}T\chi_T \left( 1 - \sum_\gamma \rho_\gamma \int {\rm d}{\bf r} \,c_\gamma({\bf r}) \right) ,
	\label{eqn:PMV}
\end{equation}
and the isothermal compressibility of bulk solvent $\chi_T$ is obtained from DRISM-KH theory.  

To mitigate systematic overestimation of hydration free energies, a universal correction is applied under ambient conditions \cite{72}:
\begin{equation}
	\mu_{\rm hydrated}^{\rm 3D-RISM-UC} = \mu_{\rm hydrated}^{\rm 3D-RISM-KH} + a \left(\rho\bar{V}\right) + b ,
	\label{eqn:mu-UC}
\end{equation}
where the regression coefficients are $a=-3.312$ kcal/mol and $b=1.152$ kcal/mol.

Furthermore, the potential of mean force (PMF) between nanoparticles can be computed by combining Lennard-Jones and electrostatic interactions with solvation free energy differences, offering a molecular-scale perspective on colloidal interactions in solution.

\subsection{Efficient numerical solution with MDIIS}

Solving the 3D-RISM-KH integral equations requires iterative numerical methods due to the complexity of three-dimensional correlation functions. The Modified Direct Inversion in the Iterative Subspace (MDIIS) algorithm \cite{73} provides a computationally efficient and stable solver with quadratic convergence properties. It reduces residuals between the integral and closure equations while minimizing memory requirements. Comparable iterative solvers include Pulay’s DIIS for Hartree-Fock equations \cite{74}, the Generalized Minimal Residual (GMRes) method \cite{75}, and Newton-Raphson approaches \cite{76,77}. The MDIIS solver has proven {to provide highly effective quasi-quadratic convergence} for large-scale nanochemical and biomolecular systems, enabling the accurate exploration of solvation structures and intermolecular forces.

\section{Effect of hemicellulose composition on nanoscale forces governing cell wall strength in decomposing resilent cell wall plant structures}

The efficient utilization of lignocellulose for producing second-generation biofuels and high-value chemicals requires the deconstruction of the highly resilient plant cell wall. The degree of recalcitrance differs significantly across plant species and phenotypes, largely due to variations in the chemical composition of the non-cellulosic matrix. Structural modifications in the composition and branching pattern of hemicellulose backbones can substantially alter the cell wall strength and microarchitecture.

Glucuronoarabinoxylan is the dominant hemicellulose in grasses widely employed for biofuel production, including sugarcane and corn. Its structure consists of a xylan backbone decorated with glucuronic acid and arabinose side chains, the distribution and frequency of which vary across plant genotypes~\cite{78}. These arabinose, glucuronic acid, and glucuronate substituents play a critical role in stabilizing primary cell walls by forming hydrogen-bonding interactions with cellulose surfaces. A molecular-level understanding of such interactions provides essential insights for genetic modification strategies and pretreatment methods aimed at enhancing biomass conversion efficiency.

To explore the role of hemicellulose in primary cell wall assembly, the three-dimensional reference interaction site model with the Kovalenko-Hirata closure (3D-RISM-KH) was employed \cite{24,25,36,61}. This statistical mechanical framework permits explicit sampling of molecular interactions, capturing both electrostatic and nonpolar contributions, including hydrogen bonding, solvophobic effects, solvent structuring, salt bridge formation, and steric constraints \cite{59}. Semi-quantitative analyses using this method confirmed the substantial contribution of hemicellulose to cell wall stabilization, highlighting how specific xylan substitutions impact the nanoscale strength. The findings further indicated that stabilization primarily originates from functional group chemistry, rather than stereochemical configurations. For example, in galactoglucomannans, monosaccharide residues such as galactose, glucose, and mannose contribute only marginally to stability, whereas carboxylate groups of randomly substituted glucuronate residues are the major stabilizing factor.

Emerging approaches to mitigate biomass recalcitrance include engineering crops to modulate lignin content and xylan substitution patterns, thereby enabling less severe pretreatment and more effective enzymatic hydrolysis \cite{13,79}. Additionally, mild lignin removal from genetically modified plants has been shown to improve cellulose accessibility without disrupting the overall wall integrity \cite{80}. Such strategies facilitate integrated biomass valorization, where hemicellulose and lignin fractions are converted into low-molecular-weight compounds, while cellulose microfibrils are retained for nanocellulose applications~\cite{81}. These approaches offer a viable alternative to complete depolymerization of lignocellulose into biofuels, which remains limited by the crystalline nature of cellulose and its resistance to enzymatic breakdown \cite{82}.

\subsection{Models of hemicellulose particles}

A primary cell wall fragment model was constructed, consisting of a cellulose nanocrystallite (CN) composed of two four-chain, eight-glucose-length cellulose fragments immersed in aqueous solutions containing varying concentrations of arabinose, glucuronic acid, and glucuronate \cite{59}. The CN models were generated using Cellulose-Builder \cite{83}.

\begin{figure}
	\centering
	\includegraphics[width=0.8\textwidth]{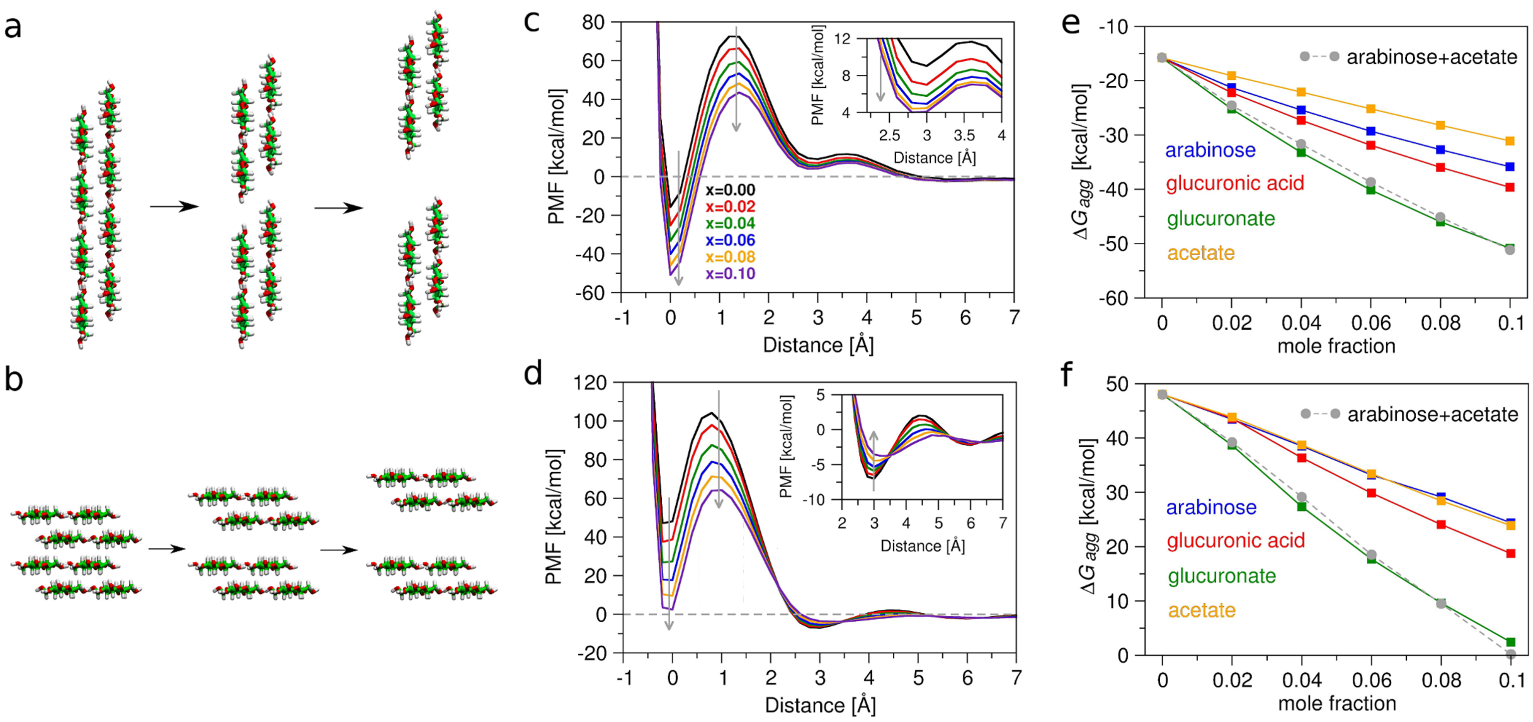}
	\caption{(Colour online) Potential of mean force $W_{\rm PMF}(d)$ and aggregation free energy $\Delta G_{\rm agg}$ \cite{59}. Disaggregation of CNs in hemicellulose hydrogels through the disruption of hydrophilic ({\bf a}) and hydrophobic ({\bf b}) interactions. PMF profiles along these disaggregation pathways ({\bf c}) and ({\bf d}) are shown for glucuronate molar fractions ranging from $x=0.0$ to $0.1$ (legend in ({\bf c}). PMF variations with glucuronate concentration are indicated by grey arrows. $G_{\rm agg}$ for hydrophilic and hydrophobic contacts in hemicellulose hydrogels is displayed in ({\bf e}) and ({\bf f}), respectively. The sum of arabinose and acetate curves is represented by a grey dotted line.}
	\label{Figure1}
\end{figure}

The 3D-RISM-KH solvation theory was applied to examine hydrophilic and hydrophobic aggregation pathways of CNs (figures \ref{Figure1}a and \ref{Figure1}b). By solving the integral equations with the CN interaction potentials, the structural and solvation free energy properties of hemicellulose around CNs were determined through three-dimensional solvent site density maps. 

The potential of mean force (PMF) between CNs in glucuronate solutions was computed as
\begin{equation}
	W_{\rm PMF}(d) = u_{12}(d) + \mu_{12}(d) + \mu_{\rm LJ}(d) - \mu_1 - \mu_2 , 
	\label{eqn:PMF-CN-CN}
\end{equation}
where $u_{12}(d)$ is the CNs interaction potential, $\mu_{12}(d)$ is the solvation free energy of the CN aggregate at separation $d$, and $\mu_1$ and $\mu_2$ are the solvation free energies of the individual CNs. The PMF curves corresponding to hydrophilic (figure \ref{Figure1}a) and hydrophobic (figure \ref{Figure1}b) contact modes are shown in figures~\ref{Figure1}c and \ref{Figure1}d.

During disaggregation, both the configurations exhibited two local minima: a direct face-to-face contact ($d_{\rm fc} \approx 0$ \r{A}) and a solvent-separated state ($d_{\rm ss} \approx 3$ \r{A}). In the hydrophilic case, the global minimum occurs at a direct contact due to extensive hydrogen bonding, whereas in the hydrophobic case the solvent-separated configuration is more stable ($\approx -7$ kcal/mol in pure water). These results suggest that CN aggregation in primary walls is predominantly mediated by hydrophilic interactions \cite{80}. Further, the outer layers of CNs are packed less densely than the core ones, and significant energy barriers must be overcome to detach glucan chains from the CN or to assemble dispersed chains into CN structures.

With a hydronium counterion modelled using \cite{84},  glucuronate increases CN aggregation by a factor of 2 compared to arabinose or glucuronic acid. The stabilization in the hydrophobic face contact (figure~\ref{Figure1}b) at 0.1 mole fraction of glucuronate results in the aggregation free energy $\Delta G_{\rm agg} \approx 0$, making aggregation spontaneous. A similar trend is observed for the hydrophilic face contact (figure~\ref{Figure1}a), where the aggregation well deepens from $\Delta G_{\rm agg} \approx -15$ kcal/mol in pure water to $-30$~kcal/mol in arabinose and $-55$ kcal/mol for glucuronate at a mole fraction of 0.1.

Increasing glucuronate concentration reduced aggregation free energy barriers while enhancing stabilization effects (figures \ref{Figure1}c--\ref{Figure1}f). Glucuronate proved to be twice as effective as arabinose or glucuronic acid in stabilizing CNs, with stabilization effects arising from both the sugar ring and the carboxylate group. Notably, aggregation free energy decreased nearly linearly with an increasing hemicellulose concentration, consistent with the displacement of water molecules from cellulose surfaces. These results also explain why alkaline pretreatments, such as sodium hydroxide, efficiently remove hemicellulose: OH$^-$ ions disrupt glucuronate-cellulose interactions, weakening the structural integrity~\cite{77}. 

The significant role of hemicellulose carboxylate groups in primary cell wall integrity poses two key questions: (i) Is the stabilization glucuronate effect due to the basic carboxylate group or the acidic hydronium ion? (ii) What is the contribution from the sugar-like cyclic part of glucuronate? The 3D-RISM-KH results with an acetate anion as a basic hemicellulose branch model and a hydronium counterion show that the effect of acetate is minor and similar to that of arabinose (figures \ref{Figure1}e and \ref{Figure1}f). About half of the effect is attributed to the sugar ring and the other half to the carboxylate group. Moreover, the sum of $\Delta G_{\rm agg}$ for acetate and arabinose is close to $\Delta G_{\rm agg}$ for glucuronate. This means that both the sugar moiety and the functional group contribute to CN stabilization (grey line in figures \ref{Figure1}e and \ref{Figure1}f).

Furthermore, $\Delta G_{\rm agg}$ decreases almost linearly with increasing hemicellulose concentration due the ability of hemicellulose branches to displace the water molecules from the cellulose surface. This agrees with previous findings \cite{13} that inhibiting the incorporation of glucuronic acid and glucuronate into xylan branches degrades the plant cell walls. The 3D-RISM-KH calculations explained why sodium hydroxide effectively removes hemicellulose from biomass {\cite{59}, in agreement with the experimental findings \cite{85}}. The strong base OH$^-$ disrupts glucuronate-cellulose interactions and destabilizes the cell wall microstructure, which can be used in the design of advanced biomass pretreatment strategies {\cite{59}}.

\subsection{Hemicellulose binding to the aggregated cellulose nanocrystallites}

\begin{figure}
  \centering
  \includegraphics[width=0.50\textwidth]{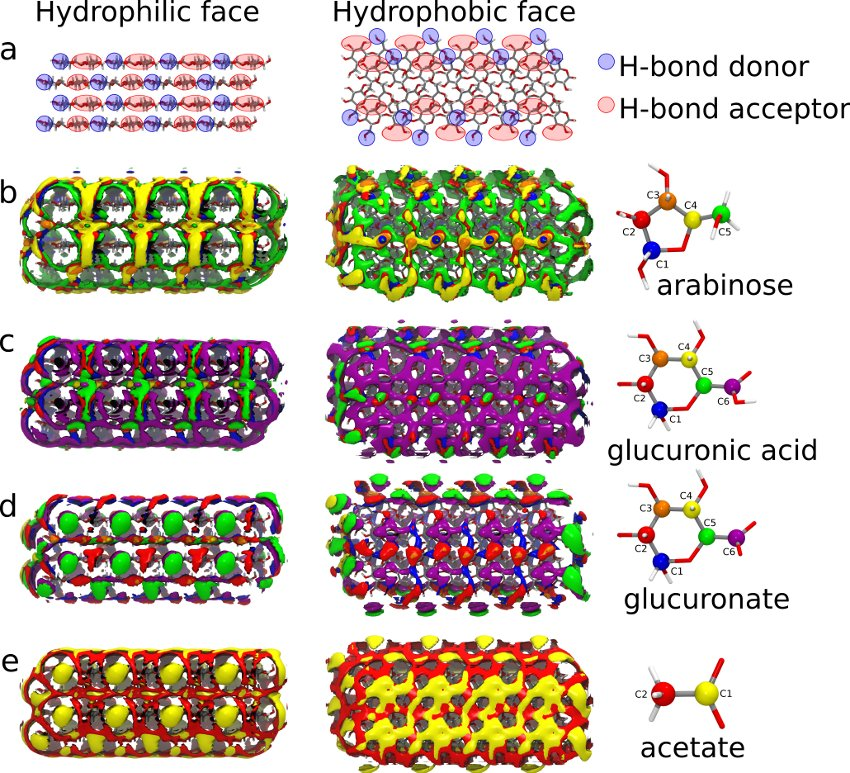}
  \caption{(Colour online) 3D site density distributions $g({\bf r})$ of aggregated CNs \cite{51}. ({\bf a}) highlights the hydrogen bond donor and acceptor sites within the CN structure. $g({\bf r})$ of hemicellulose carbon atoms from arabinose ({\bf b}), glucuronic acid ({\bf c}), glucuronate ({\bf d}), and acetate ({\bf e}) monomers in proximity to the CNs. The isosurfaces of $g({\bf r}) = 1.4$ are displayed in the same colors as the corresponding atoms, except for the glucuronate C6 atom which is represented at $g({\bf r}) = 2.0$.}
  \label{Figure2}
\end{figure}

A molecular representation of hemicellulose binding to aggregated cellulose nanostructures (CNs) is illustrated through the 3D density distributions of all carbon atoms within the corresponding hemicellulose branches, as shown in figure \ref{Figure2}. The hydrogen (H) and oxygen (O) atoms on the nanocellulose surface that participate in hydrogen bonding are identified in figure \ref{Figure2}a, distinguishing hydrogen bond donors from acceptors. Figures~\ref{Figure2}b--\ref{Figure2}e depict the regions where hemicellulose exhibits an increased local density at the nanocellulose surface relative to its bulk phase, allowing for the reconstruction of the preferred binding modes of hemicellulose branches around cellulose.

The uncharged hemicellulose branches of arabinose and glucuronic acid primarily interact with the hydrophilic faces of nanocellulose as hydrogen bond donors by coordinating with available oxygen atoms on the nanocellulose surface. Conversely, glucuronate anions act as hydrogen bond acceptors due to their strong interactions with the primary hydroxyl groups on the nanocellulose surface.

Furthermore, the distribution of acetate anions in figure \ref{Figure2}e closely resembles that of glucuronate anions, underscoring the significance of basic functional groups in coordinating with hydroxyl groups. These findings indicate that hemicellulose branches within the plant cell walls interact with the nanocellulose surface in a complementary manner, contributing to the stabilization of the cell wall microstructure. The polar interactions between the exocyclic groups of hemicellulose and the polar groups on nanocellulose dictate the preferred orientation of hemicellulose branches over the hydrophobic surface. Additionally, interactions involving endocyclic sites with the nanocellulose surface exhibit a less specific, hydrophobic character, which does not produce distinct features in the 3D site distributions.

\subsection{Solvation free energy density at the nanocellulose surface}

\begin{figure}
  \centering
  \includegraphics[width=0.55\textwidth]{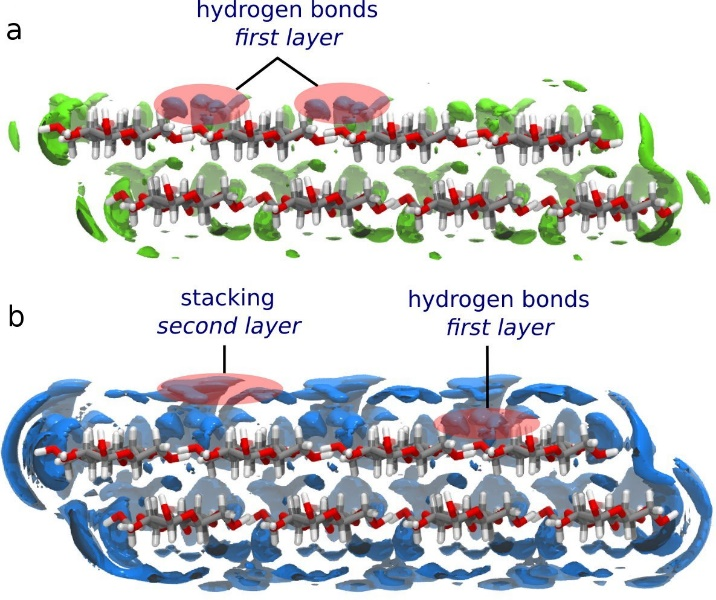}
  \caption{(Colour online) Isosurfaces of the 3D solvation free energy density of glucuronate around cellulose nanocrystals~\cite{59}. ({\bf a}) Localized regions of strong hydrogen bonding are shown in green (larger negative isovalue). ({\bf b})  Diffuse second-layer distributions with weaker interactions, shown in blue (smaller negative isovalue), represent hemicellulose–cellulose stacking.}
  \label{Figure3}
\end{figure}

Figure \ref{Figure3} illustrates the three-dimensional spatial maps of the solvation free energy density (3D-SFED) contributed by glucuronate at the nanocellulose surface. The solvation free energy denoted as $\mu_{\rm solv} = \int_s {\rm d}{\bf r} \Phi_s(\bf r)$ represents the total solvation free energy density (3D-SFED) contributions $\Phi_s(\bf r)$ from all hydrogel species $s$ integrated over the system space for CNs immersed in a hemicellulose hydrogel. These 3D-SFED maps provide a comprehensive representation of the solvation structure while encoding convolved information regarding interaction forces. They explicitly characterize the ensemble-averaged effective interactions of hemicellulose components spatially distributed over the nanocellulose surface, thereby elucidating their influence on CN aggregation.

The 3D-SFED of the glucuronate component varies from highly negative values, corresponding to its most thermodynamically favorable arrangements at the nanocellulose surface, to less negative values, indicating less favorable configurations. As shown in figure \ref{Figure3}a, the isosurface with larger negative values is highly localized around the polar sites on the nanocellulose surface, signifying hydrogen bonding interactions between hemicellulose and cellulose. By contrast, the isosurface with smaller negative values, depicted in figure \ref{Figure3}b, represents a more diffuse second layer of hemicellulose monomers stacking over the nanocellulose surface. Within this secondary layer, stacking interactions are weaker and less specific compared to hydrogen bonding, primarily arising from hydrophobic interactions and enhanced intermolecular C-H$\ldots$O interactions, as observed in the cellulose X-ray structure \cite{86}.

Although stacking interactions contribute significantly, hemicellulose-cellulose binding is predominantly driven by site-specific hydrogen bonds. The 3D-SFED maps of arabinose and glucuronic acid exhibit trends similar to those observed in the potential of the mean force $W_{\rm PMF}$ and aggregation free energy $\Delta G_{\rm agg}$ presented in figure \ref{Figure1}.

\section{Effect of lignin chemical composition on lignin-lignin and lignin–hemi\-cellulose supramolecular interactions in secondary plant cell walls}

The inherent recalcitrance of lignocellulosic biomass represents a major obstacle to the advancement of plant-based green technologies. This resistance arises from the hierarchical arrangement of cellulose, hemicellulose, and lignin within the plant cell walls, forming highly ordered and robust supramolecular structures that limit the accessibility to enzymatic hydrolysis \cite{10,11}. In particular, the three-dimensional architecture of secondary cell walls comprising cellulose microfibrils embedded within a matrix of hemicellulose and lignin operates as a protective barrier that significantly contributes to recalcitrance~\cite{10,87}.

Lignin, a heterogeneous aromatic polymer built from phenylpropanoid monomers, is a key determinant of this resistance and is considered to be the only large-scale renewable source of aromatic carbon~\cite{88,89}. Together with hemicellulose, an amorphous heteropolysaccharide \cite{90}, lignin occupies the interstitial spaces between cellulose fibrils, thereby restricting cellulose accessibility and reducing its enzymatic digestibility.

Given the complexity of lignocellulosic biomass, research has increasingly emphasized multiscale investigations into its structural and dynamic properties \cite{89}. Small-angle neutron scattering and MD simulations have demonstrated that lignin aggregates form fractal-like, multiscale structures with pores sufficiently large to accommodate cellulolytic enzymes \cite{91}. Furthermore, MD simulations indicate that lignin undergoes entropy-driven collapse under ambient conditions \cite{92}.

Studies further reveal that lignin and hemicellulose interpenetrate at their interface, underscoring a strong molecular affinity between these two polymers \cite{93}. To investigate these interactions more rigorously, the 3D-RISM-KH molecular theory of solvation has been applied, enabling the study of lignin-lignin and lignin-hemicellulose interactions as a function of lignin chemical composition \cite{60}. Unlike conventional MD simulations, this statistical mechanics based approach accounts for long-timescale solvation processes in complex aqueous environments. Results suggest that supramolecular interactions between lignin and hemicellulose play a critical role in the assembly and cohesion of secondary cell walls. These interactions are largely entropy-driven, with hydrophobic forces expelling water from polymer surfaces. Importantly, chemical composition of lignin particularly its degree of methoxy substitution-modulates both the mechanical stability and digestibility of plant biomass. Similarly, hemicellulose composition influences how effectively it integrates into the plant cell wall matrix.

\subsection{Lignin-hemicellulose and lignin-lignin interactions}

\begin{figure}
  \centering
  \includegraphics[width=0.8\textwidth]{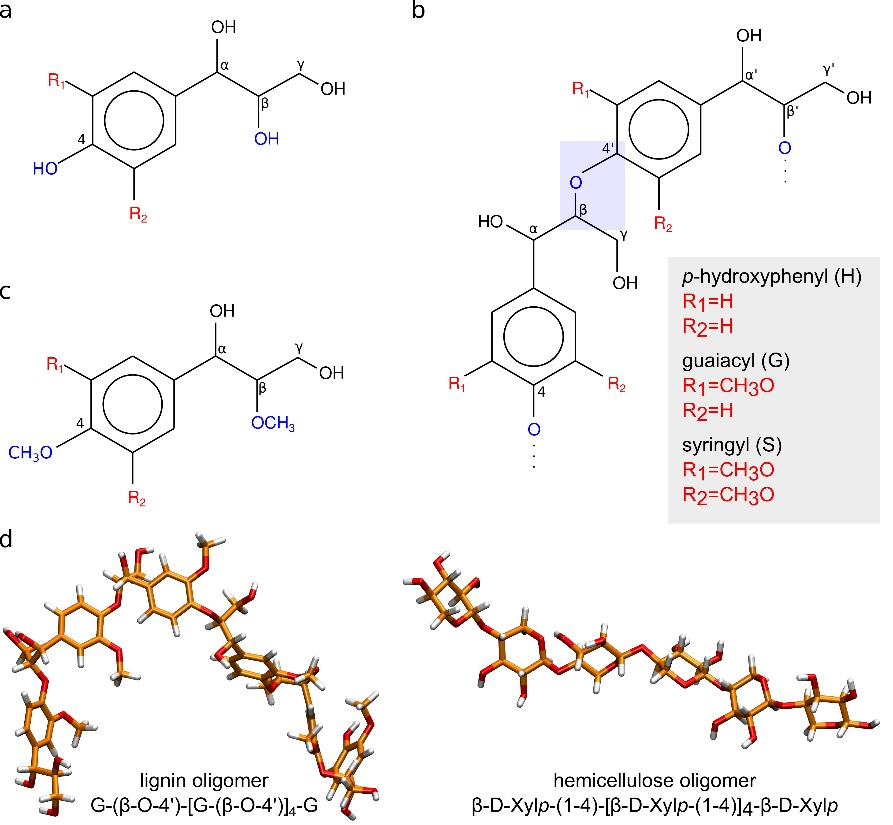}
  \caption{(Colour online) Hemicellulose and lignin oligomers \cite{60}. ({\bf a}) Lignin residues, with substituents R$_1$ and R$_2$ highlighted in red. ({\bf b}) Lignin dimer containing a $\beta$-O-4$'$ bond (light-blue background). ({\bf c}) Lignin monomer with hydroxyl groups replaced by methoxy groups. ({\bf d}) Stick model of lignin and hemicellulose oligomers, with $\beta$-O-4$'$ bonding in blue.}
  \label{Figure4}
\end{figure}

To probe lignin-hemicellulose and lignin-lignin interactions at lignin concentrations between 0 and $10^{-2}$ molar fraction, hemicellulose and lignin oligomers were modelled in aqueous solutions of the three primary lignin monomers: {\it p}-hydroxyphenyl (H), guaiacyl (G), and syringyl (S) (figure \ref{Figure4}a) \cite{60}. In lignin polymers, the dominant linkage is the $\beta$-O-4$'$ ether bond \cite{94,95}. To approximate polymeric residues more accurately (figure \ref{Figure4}b), ether linkages were substituted with methoxy groups (figure \ref{Figure4}c). Optimized monomer geometries were generated using molecular mechanics and integrated into the DRISM-KH framework \cite{60} to compute the solvation structures and thermodynamic properties.

MD simulations of 10 ns duration were performed to generate conformers of lignin and hemicellulose oligomers. Xylohexaose, composed of six $\beta \left( 1 \to 4 \right)$-linked xylose units, represented hemicellulose, while hexaguaiacyl, containing six guaiacyl residues linked by $\beta$-O-4$'$ bonds, represented lignin (figure~\ref{Figure4}d). Hemicellulose was parameterized using the CHARMM force field \cite{96}, while lignin was modelled with CHARMM parameters developed by Petridis and Smith \cite{97}. Water was represented using a modified TIP3P model \cite{98,99}. Thermodynamic properties, including solvent site-site susceptibility and solvation free energy, were evaluated using 3D-RISM-KH theory (\ref{eqn:3D-RISM})--(\ref{eqn:3D-KH-closure}) at 300 K, density $\rho = 1.0$~g/cm$^3$, and dielectric constant $\epsilonup = 78.5$. Entropy (\ref{eqn:mu-energy-entropy-b}) was calculated at temperatures 295 K and 305 K.

\subsection{Solvation free energy of the hemicellulose and lignin oligomers}

\begin{figure}
  \centering
  \includegraphics[width=0.75\textwidth]{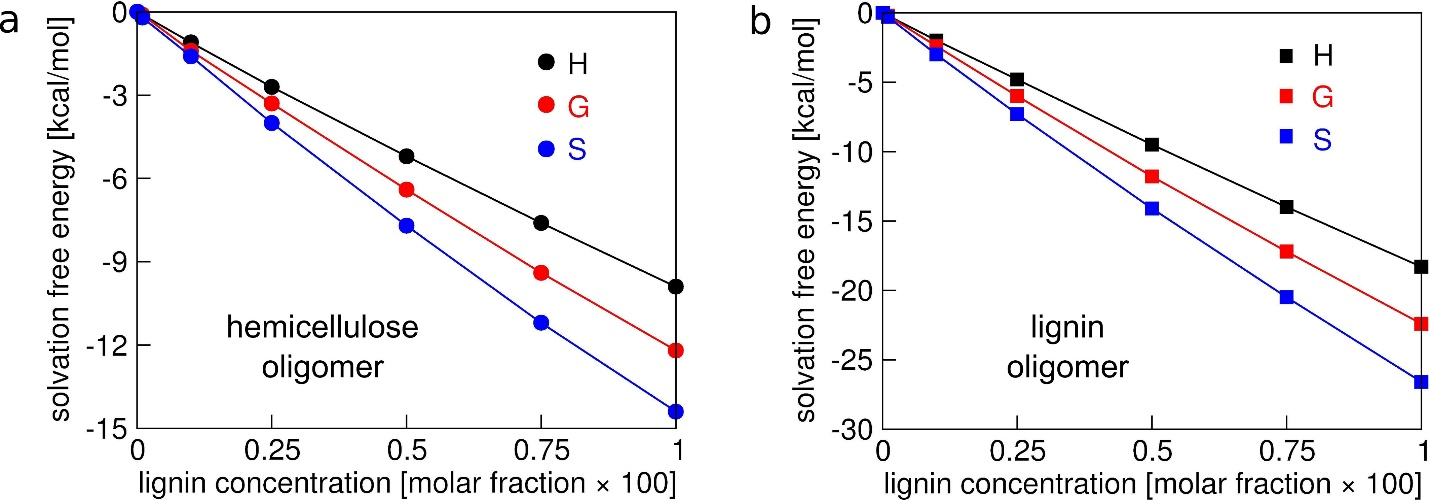}
  \caption{(Colour online) Solvation free energies of ({\bf a}) hemicellulose and ({\bf b}) lignin oligomers in aqueous solutions containing H, G, or S monomers at molar fractions $x = 0.0-0.1$, relative to pure water \cite{60}. Interaction strength increases in the order H < G < S.}
  \label{Figure5}
\end{figure}

The solvation free energies (SFEs) of hemicellulose and lignin oligomers revealed that the presence of lignin monomers stabilizes both types of oligomers, with stabilization increasing as lignin concentration rises. The strength of these interactions follows the order H < G < S, highlighting the impact of methoxy substitution on supramolecular affinity (figure \ref{Figure5}). At a molar fraction of 0.01, SFEs for hemicellulose ranged from $-10$ to $-15$~kcal/mol, while lignin values reached $-17$ to $-27$~kcal/mol depending on the monomer type. These results confirm that lignin exhibits both strong self-affinity and a pronounced tendency for aggregation.

Experimental evidence aligns with these findings: lignin composition modulates biomass digestibility, with syringyl-rich lignin imparting greater recalcitrance than guaiacyl-rich lignin, whereas p-hydroxyphenyl units are associated with reduced resistance \cite{79,100,101,102,103,104}. As the interactions mediated by S lignin are about 25\% stronger than those mediated by G lignin, a decrease in the S/G ratio reduces the secondary cell wall recalcitrance, as was experimentally observed in sugarcane and switchgrass~\cite{101,102}. Further, lignin enriched with H units was related to lower recalcitrance compared to lignin enriched with G or S units~\cite{103,104}. These experimental findings confirm the theoretical predictions \cite{60} that methoxy substituents in lignin enhance cell wall stability. Nevertheless, contradictory reports suggested that a higher S/G ratio may also enhance digestibility, underscoring the influence of genetic and environmental variability \cite{105,106}.

Importantly, lignin composition is not the sole determinant of biomass recalcitrance. Other structural factors, such as lignin–carbohydrate complexes, porosity, cellulose crystallinity, and enzyme inhibition, also contribute \cite{10,11,95}. The present solvation studies isolate supramolecular contributions, thereby providing insight into molecular-scale interactions free from larger-scale variability. The effect of lignin composition on its interaction with hemicellulose strongly affects the biomass processing, in particular, when hemicellulose removal is essential for efficient digestion, such as in switchgrass \cite{107}. The lignin composition may affect the severity of pretreatment necessary for hemicellulose extraction.

With biomass recalcitrance determined by multiple interdependent factors beyond lignin content and composition, such as biomass porosity, lignin-carbohydrate complexes, lignin-induced enzyme inhibition, and cellulose crystallinity and accessibility \cite{10,11,96}, the calculations \cite{60} illustrated the effects of supramolecular interactions in the plant cell walls while excluding other uncontrolled factors at larger scales.

\begin{figure}
  \centering
  \includegraphics[width=0.8\textwidth]{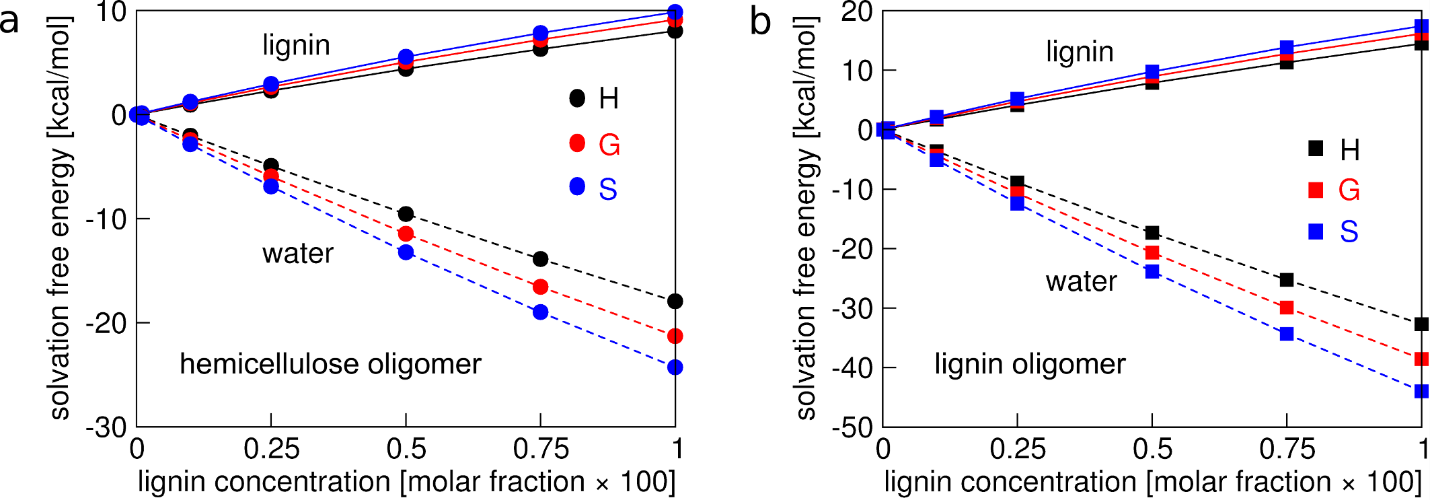}
  \caption{(Colour online) Contributions of lignin and water to solvation free energy \cite{60} for hemicellulose ({\bf a}) and lignin ({\bf b}) oligomers as a function of lignin concentration for p-hydroxyphenyl (H), guaiacyl (G), and syringyl (S) lignin monomers. Solid lines denote water contributions, dashed lines represent lignin contributions. Total SFE values correspond to the sum of these contributions (figure~\ref{Figure5}).}
  \label{Figure6}
\end{figure}

At the molecular level, hemicellulose-lignin and lignin-lignin interactions display similar characteristics, since changes in lignin concentration or chemical structure yield comparable effects on the surface free energy (SFE) of both oligomers. Figure \ref{Figure6} shows the respective water and lignin contributions to SFE, derived by partitioning the solvation free energy~(\ref{eqn:mu-excess}) into interaction components.

For hemicellulose (figure~\ref{Figure6}a), the favorable negative contribution from water increases with lignin concentration, while the lignin contribution increases unfavorably. Entropy loss from restricted lignin monomer motion is only partly offset by lignin interactions, making the water contribution the dominant stabilizing factor. The effect of lignin composition is stronger in the water contribution, following H < G < S, consistent with the overall SFE trend. A similar mechanism drives lignin-lignin interactions, with methoxy substitution further enhancing water contributions (figure~\ref{Figure6}b).

SFE and solvation entropy values at 300 K (table~\ref{Table1}) confirm that entropy increases with methoxy content (H < G < S), explaining the observed stabilization of both hemicellulose-lignin and lignin-lignin interactions. These results indicate predominantly hydrophobic interactions, where dehydration of interfaces facilitates favorable water contributions. Increased methoxy substitution strengthens the lignin hydrophobicity, enhancing solvation entropy and water exclusion.

\begin{figure}
  \centering
  \includegraphics[width=0.65\textwidth]{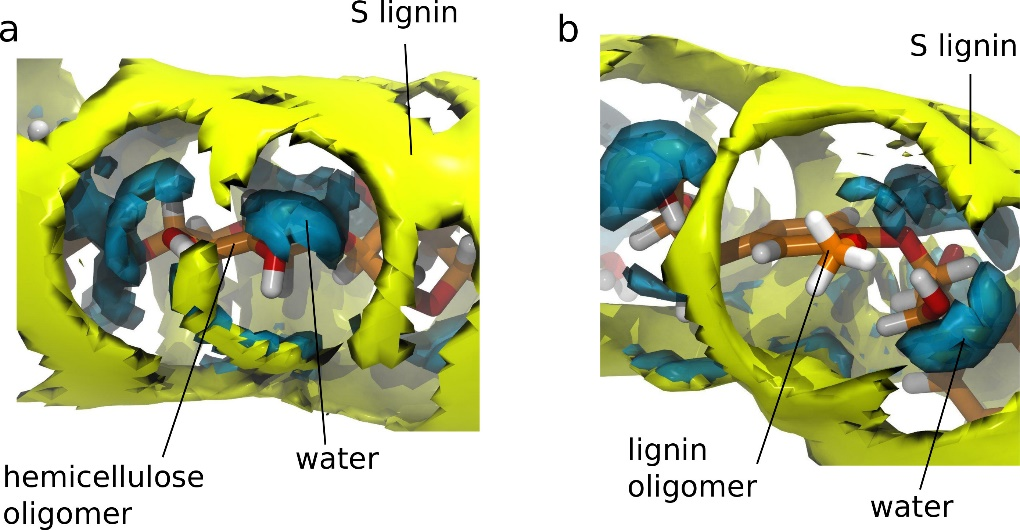}
  \caption{(Colour online) 3D-SFED isosurfaces of a water molecule (cyan) and a syringyl lignin monomer (yellow) around hemicellulose ({\bf a}) and lignin ({\bf b}) oligomers, shown in stick form \cite{60}. Water isosurfaces highlight hydrogen bonding with polar groups, while syringyl distributions reflect non-specific hydrophobic stacking in hemicellulose-lignin and lignin-lignin interactions.}
  \label{Figure7}
\end{figure}

Relative hydrophobicity is also supported by DRISM-KH predictions of bulk compressibility, which increases with lignin concentration and composition (H < G < S; figure \ref{Figure7}). These results align with MD simulations showing the entropy-driven lignin collapse and solvent-mediated lignin-cellulose association at ambient conditions. Overall, lignin probably contributes to biomass recalcitrance through its hydrophobic interactions with cellulose, emphasizing its role in the plant cell wall structure and stability.

\renewcommand{\arraystretch}{1.5}

\begin{table}
\begin{center}
\caption{Solvation free energy $\Delta\mu$ and solvation entropy $T\Delta S_V$ of hemicellulose and lignin in an aqueous solution containing lignin monomers at a molar fraction of $x = 0.01$ and a temperature of 300 K, expressed relative to H lignin \cite{60}.}
\begin{tabular}{ c|c|c|c|c } 
  \hline
    z  &  \multicolumn{2}{ |c }{ Hc oligomer }  &  \multicolumn{2}{ |c }{ Lignin oligomer }  \\
  \hline
  Monomer  &  $\Delta\mu$ [kcal/mol]  &  $T\Delta S_V$ [kcal/mol]  &    $\Delta\mu$ [kcal/mol]  &  $T\Delta S_V$ [kcal/mol]  \\     
  \hline
  H   &	   0   &   0   &    0   &   0   \\
  G   &  $-2.2$  &  2.2  &  $-4.1$  &  5.0  \\
  S	  &  $-4.5$  &  3.8  &   8.3  &  6.8  \\
  \hline
\end{tabular}
  \label{Table1}
\end{center}
\end{table}

\subsection{Solvation free energy density of hemicellulose and lignin oigomers in solution}

The three-dimensional solvation free energy density (3D-SFED), calculated using equation~(\ref{eqn:mu-excess-b}), is illustrated in figure~\ref{Figure7}. These isosurfaces provide a molecular-scale view of the supramolecular interactions underlying the thermodynamic properties described in earlier sections. The isovalues corresponding to the complete first solvation shell of water remain unchanged, while larger isovalues capture the second hydration layer. By partitioning the 3D-SFED contributions from water and different lignin monomers, a molecular-level representation of the effective interactions was achieved; namely, those between hemicellulose and lignin oligomers (solute) and between water and lignin monomers (solvent).

This analysis highlights the supramolecular interactions between lignin and hemicellulose, the two predominant heteropolymers of the non-cellulosic matrix in the secondary plant cell walls, while also assessing how variations in lignin composition affect their interaction strength. Both hemicellulose-lignin and lignin-lignin interactions display comparable features, dominated by hydrophobic and non-specific associations. Such interactions are probably critical to the organization of the amorphous secondary wall, potentially driven by entropic forces that limit water accessibility to hydrophobic domains. Furthermore, the chemical structure of lignin, particularly the degree of methoxy substitution, significantly influences the cohesive interactions within the wall, thereby shaping its mechanical resilience. Studies employing model compounds rather than intact plant cell walls offer an effective means of probing the molecular mechanisms underpinning lignocellulosic biomass recalcitrance.

\section{Conclusion}

Cellulose, the most abundant biopolymer on Earth, serves as a vital renewable resource for sustainable biofuel production. The susceptibility of lignocellulosic biomass to depolymerization can be enhanced through enzymatic or chemical pretreatments that disrupt cellulose microfibrils, often within thermochemical processes. Solvents play a central role in these treatments, and elucidating the molecular mechanisms underlying biomass modification is critical for improving their efficiency.

In this context, the three-dimensional reference interaction site model with the Kovalenko–Hirata closure (3D-RISM-KH) molecular solvation theory was applied to investigate the interactions of cellulose nanocrystals embedded in a hemicellulose hydrogel. The analysis predicted how variations in hemicellulose chemical composition influence the nanoscale forces that regulate the assembly of primary cell walls. Furthermore, the study demonstrated that lignin-lignin and lignin-hemicellulose associations are predominantly hydrophobic and entropy-driven, arising from the displacement of water molecules at their contact surfaces.

These findings provide valuable insight into the thermodynamic principles governing the biomass restructuring, particularly during high-temperature treatments that increase the average diameter of cellulose microfibrils. By advancing the molecular-level understanding of biomass pretreatment, the study contributes to the development of more effective strategies for plant biomass conversion. Additionally, the results support integrated valorization approaches, in which non-cellulosic fractions are depolymerized into low-molecular-weight chemicals, while cellulose microfibrils are utilized for the production of nanocrystalline and nanofibrillar cellulose.

\section{Funding}

This work was supported by the Natural Sciences and Engineering Research Council of Canada (Research Grant RES0029477), Alberta Prion Research Institute Explorations VII (Research Grant RES0039402), and Alberta Innovates--Bio Solutions (Research Grant RES0023395). The computations were supported by WestGrid--Compute/Calcul Canada.

%{\small \topsep 0.6ex

%}
\ukrainianpart

\title[Теорія геміцелюлози та лігніну в розчині]%
{Багатомасштабна теорія та моделювання геміцелюлози і лігніну в розчині}

\author[A.~Коваленко]{A.~Коваленко}
\address{Центр програмного забезпечення для багатомасштабного моделювання, Едмонтон, Альберта, Канада, T6E 5J5 }

\makeukrtitle

\begin{abstract}
	\tolerance=3000%
	У цьому огляді розглядаються підходи до багатомасштабного моделювання целюлозних нанокристалів (ЦНК) та лігноцелюлозних клітинних стінок рослин з акцентом на взаємодії геміцелюлози та лігніну у водних середовищах. Тривимірна модель опорного сайту взаємодії із замиканням Коваленко-Хірати (3D-RISM-KH) лежить в основі перспективної теорії молекулярної сольватації, що застосовується в нанохімії та біомолекулярному моделюванні. Запропонований метод був успішно застосований для дослідження геміцелюлозних гідрогелів, впливу складу геміцелюлози на нанорозмірні сили в первинних клітинних стінках, а також взаємодій лігнін-лігнін та лігнін-геміцелюлоза. Результати дослідження показують, що ці взаємодії є переважно гідрофобними та зумовленими ентропією і виникають внаслідок ефектів виключення води. Результати моделювання, отримані за допомогою цієї моделі, поглиблюють розуміння молекулярних сил у клітинних стінках рослин та допомагають розробити стратегії утилізації біомаси, включаючи генну інженерію та технології попередньої обробки, спрямовані на покращення екстракції та використання целюлози.

	\keywords нанокристали целюлози, геміцелюлоза, лігнін, теорія молекулярної сольватації, термодинаміка сольватації, молекулярне моделювання
\end{abstract}
\end{document}